# International Research Collaboration:
# Novelty, Conventionality, and Atypicality in Knowledge Recombination

**Authors:** Caroline S. Wagner[1], Travis A. Whetsell[2]*, Satyam Mukherjee[3]

**Abstract:** Research articles produced through international collaboration are more highly cited than other work, but are they also more novel? Using measures developed by Uzzi et al. (2013), and replicated by Boyack and Klavans (2014), this article tests for novelty and conventionality in international research collaboration. Scholars have found that coauthored articles are more novel and have suggested that diverse groups have a greater chance of producing creative work. As such, we expected to find that international collaboration tends to produce more novel research. Using data from Web of Science and Scopus in 2005, we failed to show that international collaboration tends to produce more novel articles. In fact, international collaboration appears to produce less novel and more conventional knowledge combinations. Transaction costs and communication barriers to international collaboration may suppress novelty. Higher citations to international work may be explained by an audience effect, where more authors from more countries results in greater access to a larger citing community. The findings are consistent with explanations of growth in international collaboration that posit a social dynamic of preferential attachment based upon reputation.

[1] John Glenn College of Public Affairs, The Ohio State University, Columbus, OH USA 43210
Wagner.911@OSU.edu

[2] Department of Public Policy & Administration, Steven J. Green School of International and Public Affairs, Florida International University, Miami, FL, USA, 33155
Travis.Whetsell@FIU.edu *Corresponding author

[3] Department of Operations Management, Quantitative Methods and Information Systems, Indian Institute of Management, Udaipur, India
Satyam.Mukherjee@gmail.com

## 1 - Introduction

International collaboration in scholarly research, resulting in published articles, has risen at a spectacular rate for three decades. The phenomenon has attracted a good deal of attention, focusing inductively on measurement rather than providing theoretical explanations. Literature shows that internationally coauthored articles tend to be more highly cited than national coauthorships or sole authored work (Glänzel & Schubert, 2001; Glänzel and deLange, 2002); that the more elite the scholar, the more likely it is that they are working at the international level (Jones et al., 2008; Parker et al., 2010); and that it tends to be more interdisciplinary (van Raan, 2003), suggesting new combinations. We do not know if international research is more novel than domestic or sole-authored research. This article explores the gap in the literature by applying measures of novelty and conventionality to international collaborative publications.

Creativity and novelty are universally valued in scholarship, but this value is difficult to measure. The team of Brian Uzzi, Satyam Mukherjee, Mark Stringer, and Ben Jones, hypothesized that truly novel advances are accompanied by strength in conventional know-how, joined to a novel idea, often resulting from unique or atypical combinations of prior knowledge (Uzzi et al., 2013). Drawing upon the tradition where the reference to preceding work serves as an elementary building block of the scholarly attribution and reward system, they used this artifact as a proxy for novelty. Using this model of novelty and conventionality, they tested whether referencing behavior with deep conventionality combined with novelty is also more highly cited than just conventional or just novel work. The analysis of more than 17 million articles from the Web of Science (WoS) database supported the hypothesis.

We follow this premise about the value of examining combinations of novelty and conventionality at the level of the full database. Further, we add to the original Uzzi et al. (2013) WoS data, and include data from a replication study using Scopus, to examine the validity of the approach. We examine the relationship between atypical combinations of conventionality and novelty and internationally coauthored work with the expectation that international collaboration would also be more likely to produce atypical combinations. This article presents the results of this test.

The article is organized as follows: First, we review the literature on international collaboration, atypicality, and creativity. Next, we present the data, variables, and methods for this study. We then present the results of our analysis by examining international collaboration in light of the conventionality/novelty data. We further explore the data by testing for differences between natural science and engineering, social sciences, and the arts and humanities, as well as at the level of six disciplines. Finally, we discuss the findings and their implications for future research and for public policy.

## 2 – Literature Review

### 2.1 Growth of International Collaboration in Science

International collaboration is defined here as occurring when researchers from two or more nations list their names and addresses as authors on a scholarly article. We do not know if the coauthors operated as teams, collaborations, or inter-institutional cooperators (Katz & Martin, 1997). However, we dub these artifacts to be 'collaborations,' assuming that they reflect mutual intellectual and social influence. International collaboration has grown more than 10-fold since 1991 for the most advanced countries, and 20-fold for the BRICs (Brazil, Russia, India, and China) (Adams, 2013). Many more nations participate in these publication activities than was the case two decades ago (Bornmann et al., 2015).

In Web of Science (WoS) the percentage of articles, notes, and letters that are internationally coauthored has grown from 10% in 1990 to 25% in 2011(Wagner et al. 2015). The number of authors per article has also grown. Wuchty et al. (2007) analyzed what they called "team size," showing growth in numbers of coauthors across all fields of scholarship (national and international) in the Web of Science from over a 45-year period. Gazni et al. (2012) supported Glänzel and de Lange (2002) showing growth in the number of coauthorships per article at the international level. Using SCImago Institutions Rankings, Benavent-Perez et al. (2012) show decreases in sole-authored publications. Figure 1 shows the changes by selected regions and countries, where the United States, the European Union, and Japan have produced steadily increasing international research, while China, Taiwan, and South Korea have lagged behind.

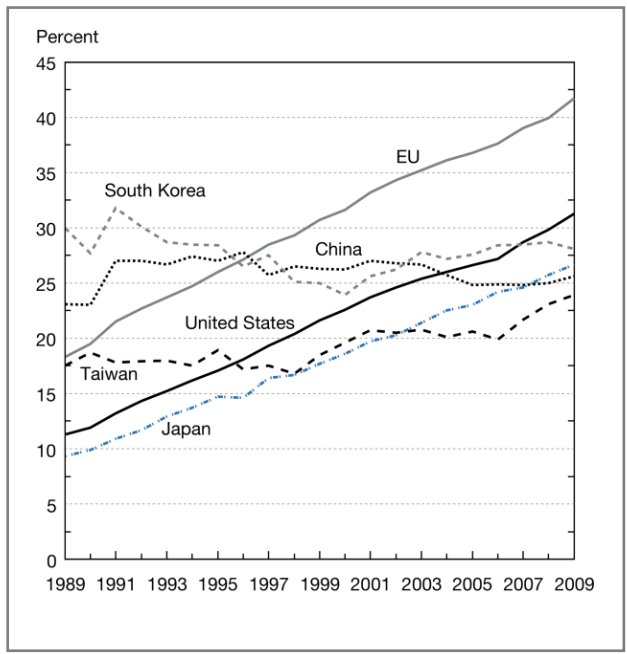

Figure 1 – Percent of Research Articles with International Coauthorships by Major Scientific Contributors, 1989-2009, Source: National Science Board, 2012

International works are more likely to be highly cited (Smart & Bayer, 1986; Glänzel and Schubert, 2001; Persson et al., 2004; He, 2009; Ganzi et al., 2012). Narin et al. (1991) showed that internationally coauthored articles were cited two times the rate of ones from a single country. Wagner et al. (2017) examined international collaborations in six scientific disciples showing that a rise in the number of countries-per-article was also linked to higher citations. The number of countries per article is also shown to be significant for increased citations (Didegah & Thelwall, 2013; Wagner et al., 2017). Glänzel (2001) showed that international publications have higher-than-expected citation rates in all scientific fields. Glänzel and de Lange (2002) showed growth in the number of authors per article at the international level, as well as growth in the average number of organizations and nations per co-publication. Internationally collaborative research seems to be attracting greater attention than domestically produced research (Lariviere et al., 2015).

Lancho-Barrantes et al. (2013) and Archambault et al. (2016) support Presser (1980) in finding that international collaboration is associated with greater scientific quality. Lariviere et al. (2015) showed that multiple-countries per article are more likely to have atypical combinations of referenced work than single or pair-authored articles. Others have pointed to improved performance of international collaborations (Fleming, 2007). More heterogeneous teams or collaborations, which we assume is the case with international projects, have greater opportunity to leverage the expertise of members and to bring a wider range of information to the knowledge creation process (Bercovitz & Feldman, 2011; Cohen & Levinthal, 1990).

## 2.2 - Atypical combinations reflecting novel contributions

Originality is a core goal of scholarship, but scholars must show mastery of a field when introducing new findings. A fine balance must be struck between demonstrating depth and mastery of a discipline and bringing original ideas. Depth is required to demonstrate *bona fides*, signaling to in-groups, using common language, accepted methods, and citations to the work of others. The "new" contributions, ones that bring original ideas, often involve incorporating ideas from different disciplines, research traditions, or frameworks[4]. Novel claims are validated based upon strong evidence, but in cases where new ideas are incorporated across disciplines, the fresh concepts can take longer to assimilate because they do not fit the conventional narrative, language, or culture (Wang et al., 2016; Cetina, 2009). This can make it risky for researchers to introduce truly novel ideas (Stephan et al., 2017; Simonton, 2004), especially without tying them to precedents.

In discussing the intellectual and social organization of the sciences, Whitley (1984) suggested that disciplines are "systems of jointly controlled novelty production in which researchers have to make new contributions to knowledge in order to acquire reputations from particular groups of colleagues…" (p. 85). Reputation is often the result of producing scholarly work that offers a novel discovery, new method, advanced theorem, or insight—something that solves a puzzle or gives a new perspective on one. Stephan (2012) follows Merton (1973) defining reputation as "being built by being the first to communicate a finding, thereby establishing priority of discovery…." (p. 5).

The scholarly article is the venue to communicate findings. Price (1965) and Merton (1973) regarded the publication of a scientific article as acknowledgment of its original contribution.

---

[4] Combinatorics is distinct from *discovery*, where a new object, such as an exoplanet or a bacteria, may be found but no new combinations of knowledge are indicated in the references of publications announcing it.

Articles contain lists of references, citing prior work; by doing so, scholars reconstruct the intellectual antecedents and sources of their knowledge claims (Fujigaki, 1998). In theory, an article that has passed peer review should be adding something new to a field. However, more than 50 percent of published articles are not cited at all, suggesting that some material does not capture attention (Hamilton, 1990). Whether the contribution is acknowledged is often a social outcome with acquaintances, colleagues, and co-nationals citing one another for social, as well as research-related reasons (Bornmann & Daniel, 2008; Garfield & Merton, 1979). Highly cited work contributes to setting the agenda for further research. This part of the process may require a communication strategy, combined with reputation, as much as it does the introduction of new ideas. Whether a finding is truly novel--in the sense of highly creative, risky, or revolutionary-- often cannot be determined at the time of publication; instead, it may require some length of time for the scholarly community to recognize the validity of a new finding, or time for promoters of the ideas to disseminate them. Examining citations over time can provide some insight into the variety of impacts among scholarly publications (van Raan, 2014; Bornmann & Daniel, 2008).

The need to publish and gain attention to one's work means that there are few high-risk articles. The ones who take risks may be well positioned to gain attention, but "the additional reward does not compensate for the risk of failing to publish…." (Foster et al., 2015, p. 875). Similarly, Estes & Ward (2002) said "creative ideas are often the result of attempting to determine how two otherwise separate concepts may be understood together…" (p. 149), which can represent new ideas, but which is a risky strategy, especially if the separate concepts are from different disciplines. Interdisciplinary articles may fail peer review: "…creative thought calls for the combination and reorganization of extant categories…." (Mobley et al., 1992, p. 128). However, they also point out the higher risk of failure (see also Stephan et al., (2017)).

There are no specific signals for identifying the originality, novelty, or creativity likely to produce a highly cited article. Rather, decisions about the potential contribution of new research are likely to be made as judgments by editors and reviewers during the peer-review process. If an article does not introduce a new idea in a way that satisfies the reviewers or journal editors, the work may be criticized as "too conventional", i.e. not adding something new to the literature. On the other hand, if an article is too novel, radical, or revolutionary it may not pass muster with reviewers and editors.

One way to measure this combinatorial dynamic is after publication by searching lists of references for unexpected combinations. The resulting atypical referencing behavior serves as a proxy measure for recombination of knowledge. Börner et al. (2012) visually maps fields of science, showing the distance between them and just how unlikely it might be to bridge gaps between, for example, fields such as brain research and ecology. These combinations can be indicated in diverse reference pairs (See Schilling & Green, 2011 for an excellent literature review). To explore this, Uzzi et al. (2013) hypothesized that the highest impact articles are likely to reference novel combinations of existing knowledge, which remain embedded in and supported by deeply conventional knowledge. In other words, novel findings will emerge from depth of understanding of supporting knowledge, joined to the new idea. Further, variety and cross-cultural diversity ("cognitive distance") could be contributing to greater creativity and therefore drawing attention to this work (Lee et al., 2015; Alshebli et al., 2018). The reach of an international network might extend the "search space" of a group and thereby result in access to more new ideas (Schilling and Green 2015). Larivière et al. (2015) examined 9.2 million interdisciplinary research articles published between 2000 and 2012 and showed that the majority (69.9%) of referenced interdisciplinary pairs resulted in "win-win" relationships, i.e., articles that cite these diverse articles have higher citation impact. Articles with "atypical combinations"—statistically rare journal-journal combinations--attracted the highest relative number of citations.

We expect those participating in international collaboration to have enhanced access to a diversity of ideas. The cultural differences brought by researchers from different countries may enrich idea generation. Schilling & Green (2015) note: "The work on recombinant search lends support to this position by noting that unfamiliar or atypical combinations of knowledge yield novel outcomes with greater variance in performance…" (p. 1320). They further note that "…[k]nowledge creation occurs when new information is integrated within the network, or when the existing information within the network is recombined in new ways…" (p. 1323), conditions that are more likely to be found in international groups. These insights lead to the question of how researchers might measure creativity and novelty to produce general insights relevant to science policy.  This finding motivated our research.

*2.3 - Creativity and novelty in scholarly publications*

There are many different definitions of creativity and novelty in the literature, and the literature is quite extensive. It is beyond this paper to survey the entire field. This paper focuses on those efforts to measure the phenomenon. Directly relevant literature suggests that creativity and novelty are likely to emerge from diverse teams of people (Falk-Krzesinski et al., 2010; Fiore, 2008; Stokols et al., 2008; Wuchty et al., 2007; Uzzi & Spiro, 2005). Mâsse et al. (2008) suggest that groups of researchers have a greater chance of creating novel findings because they are "integrating the concepts and methods drawn from multiples disciplines and analytic levels…" (p.S152). Estes & Ward (2002) discuss the literature around creative functioning of groups, noting that "a common thread connecting these views is that creative ideas are often the result of attempting to determine how two otherwise separate concepts may be understood together…." (p. 149). Schilling & Green (2011), citing Simonton (1995, 1999), note that "…research suggests that ideas are more likely to be high impact when they are the result of a successful connection forged between seemingly disparate bodies of knowledge…" (p. 1321). Guimera et al. (2005) suggest "creativity is spurred when proven innovations in one domain are introduced into a new domain, solving old problems and inspiring fresh thinking…" (p.697).

Scientific achievements are often described as a search in space of combinatorial possibilities, leading to fresh insights and technological breakthroughs. Prior works on technology, history, anthropology, and archaeology have addressed the key role of "combinations" in linking innovation and scientific and technological impact. Azoulay et al. (2011) note that most efforts to model the creative process consider it as cumulative, interactive recombinations of existing bits of knowledge, combined in novel ways. In their review seeking novelty in patent data, Youn et al. (2015) cited earlier work that "…posits the combination of new and existing technological capabilities as the principal source of technological novelty and invention…" (p.1).

Foster et al. (2015) noted that "[a]n innovative publication is more likely to achieve high impact than a conservative one…" (p. 875) but these actions fail more frequently than they succeed. Stephan et al. (2017) found that high risk articles are more likely to be in the top 1% highly cited articles, although recognition can be delayed (Wang et al, 2017). Wang et al. (2017) found that highly novel articles are more likely to be found in journals with lower Impact Factors— suggesting that atypical combinations may have a difficult time finding acceptance in the canon of literature. Schilling & Green (2011) note that atypical variance can lower average performance of outcomes, but they also can contribute to exceptionally high-performing outcomes.

The Uzzi et al. (2013) team showed that works with a combination of high novelty and high conventionality were twice as likely as the average article to be 'hit' articles. For their analysis, the team used Web of Science's 50-year data set of articles to analyze journal-journal references. They sought to determine whether pairs of cited references are conventional (more than expected by chance) or novel (less than expected by chance). Using cited references from nearly 18 million articles, they calculated actual and expected counts for 302 million references to 15,613 journals. For each referenced journal pair, these were converted into z-scores. Ten Monte-Carlo simulations were run that reassigned edges in random ways, while preserving temporal and distributional characteristics of the original citation network.

Boyack and Klavans (2014) conducted a complementary analysis to replicate the Uzzi et al. (2013) findings, this time using Scopus 10-year data set of articles and conference papers. They used a K50 statistics method for journal pairs rather than using the z-scores and Monte Carlo simulations. They claim that K50 has the same general formulation as the ones used by Uzzi et al. (2013). The difference is that the expected and normalization values for K50 are calculated using row and column sums from the square co-citation count matrix rather than using a Monte Carlo technique. Boyack and Klavans (2014) were able to reproduce the Uzzi et al. (2013) findings for z-score distributions and citations to a high degree, despite using a different database. However, they found a disciplinary effect (discussed below).

Using a similar approach, Stephan, Veugelers & Wang (2017) and Wang, Veugelers & Stephan (2017) also tested published articles for atypicality. They drew from WoS for the year 2001 and used a "commonality score" which measured the expected number of co-citations. This is measured as a joint-probability of co-occurrence of two journals in a reference list, times the number of all journal pairs. The difference between Stephan et al. (2017) and Uzzi et al. (2013) is in the novelty score: Uzzi et al. (2013) measured the z-scores using a null model. Stephan et al. (2017) used a ratio of observed over expected, where 'expected' is analytically estimated based upon data in the analysis. Further, Stephan et al. (2017) examined a second-order co-citation distance based on previous years' co-citations, or, journal pairs that were not co-cited before (similarity equals to 0 based on first-order co-citations), and then they examine similarity based on co-cited journals (that is, measuring their similarity using second order co-citations). They found that more novel articles were more likely to be either a big hit or ignored, supporting Foster et al. (2015) and partially supporting Uzzi et al. (2013).

These articles have explored the effects of novelty and conventionality on citation impact, providing evidence that atypical combinations of novelty and conventionality have the highest probability of producing a 'hit' article. In this article, we are more concerned with identifying whether international collaboration tends to produce, novel, conventional, or atypical referencing behavior. This article is more focused on the antecedents of novelty, conventionality, and atypicality, than the citation impact of international collaboration. The following section describes the data, variables, and methods used in this article.

### 3 - Data, Variables, and Methods

Our primary data set is from the Uzzi et al. (2013) study, which contains roughly 850,000 articles from Web of Science in the year 2005 and includes metadata on authors, affiliations, fields, references, and citations. For robustness checks, we added four years of WoS data (2001-2004), which together with 2005 data comprised roughly four million observations. We also used 2005 Scopus data, employed by Boyack and Klavans (2014), which included roughly one million observations (see Section 4.1 and Appendix for robustness checks).

We applied the Uzzi et al. (2013) method of identifying conventionality and novelty to test whether internationally collaborative articles fall into the 'atypical' (high-novel/high-conventional) quadrant. The computation analyzed pairwise combinations of references in the bibliography of each article from the Web of Science for the 5 years studied. The counts measured the frequency of each referenced journal pair across all articles in the database (not disciplinary specific) and compared the observed frequency to those expected by chance. Z-scores for each journal pair were derived by comparing the observed frequency with the frequency distribution created with the randomized citation network. This procedure resulted in two summary statistics: the $10^{th}$ percentile z-score of a article, and the median $z$-score of that article. While the former quantifies the novelty of an article, the latter tells us about the central tendency of the journal combinations of the article (Mukherjee et al., 2015).

In Figure 2 we illustrate the probability distribution of 10[th] percentile z-scores of all articles published in 2005, as indexed in WoS database. Using the approach from Uzzi et al. (2013), a journal pair with 10[th] percentile z-score less than 0 signifies a "novel combination". Note, for the analysis where 10[th] percentile z-score is used as a continuous measure, the variable was reverse coded for consistent interpretation as "novelty".

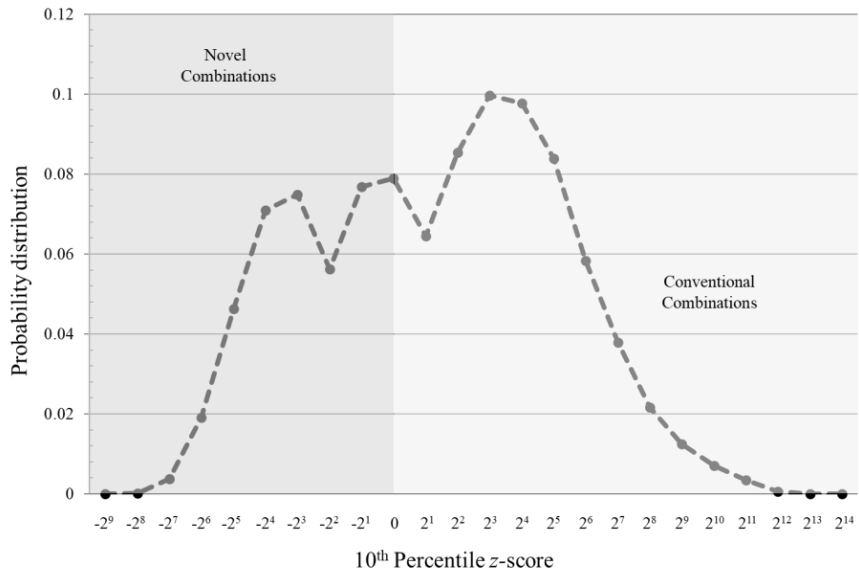

Figure 2. Probability Distribution of 10[th] percentile z-score, 2005. The figure shows the distribution of z-scores where values below zero are considered novel, else conventional.

Figure 3 shows the probability distribution of median *z*-score of all articles published in 2005. Since a *z*-score value more than 0 signifies an observed journal pair frequency appears more than what is expected by chance, we call these "conventional combinations" (Mukherjee et al., 2015).

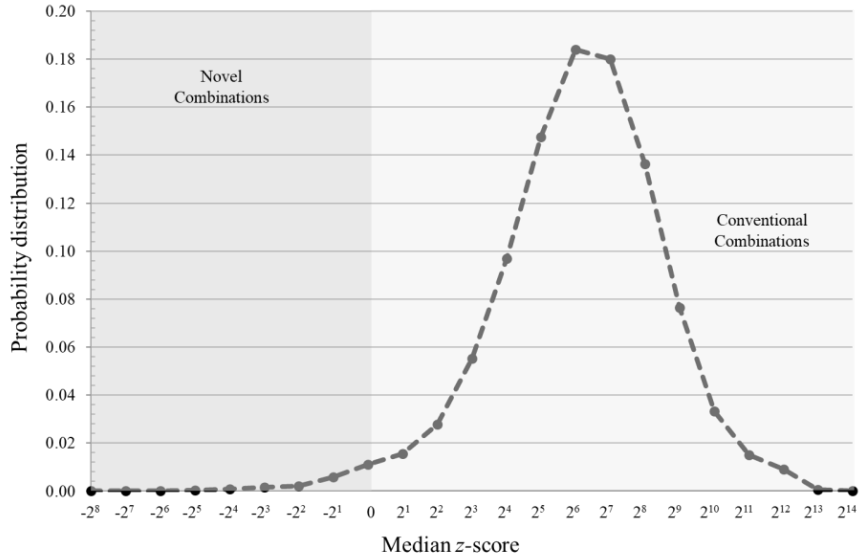

Figure 3. Probability distribution of median z-score, 2005. The Figure shows values below zero are considered more novel. For the analysis, we consider values above the grand median are highly conventional.

In keeping with the procedure specified in Uzzi et al. (2013), these continuous variables are binarized and then combined into a nominal variable. For novelty, 10th percentile z-score values below zero are coded as 1, representing high novelty, or HN; values above or equal to zero are coded as 0, representing low novelty, or LN. For conventionality, median z-score value above or equal to the overall median is coded as 1 for high conventionality, or HC; and below the overall median is coded as a 0, for low conventionality, or LC. We combine two binary variables into a nominal variable, illustrated in Figure 4, where Category 1 is high novel and high conventional, or HN/HC; Category 2 is high novel and low conventional, or HN/LC; Category 3 is low novel and high conventional, or LN/HC; and Category 4 is low novel and low conventional, or LN/LC. Figure 4 also shows the frequency of articles that belong to each category. Note that the categories in Figure 4 are employed in the multinomial logistic regression models in Tables 4 and A.6.

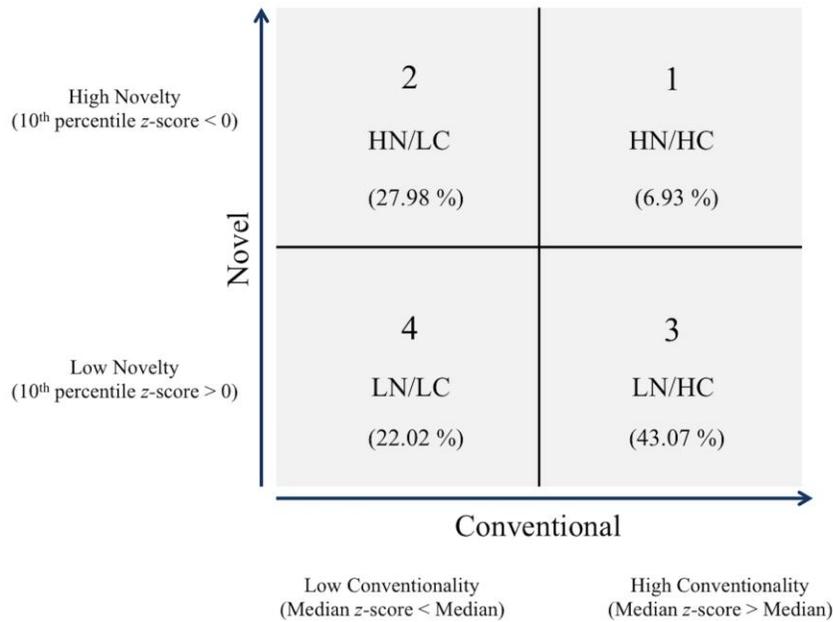

Figure 4. Nominal variable categorizing articles based on a combination of novel and conventional reference combinations. The numbers in the parentheses provide the percentage of papers in each category.

Finally, we included control variables for 1) number of authors per article (Authors), 2) number of references per article (References), and 3) a subject category variable including 238 WoS categories (Field). For robustness checks we also made use of log transformations of countries and authors, as well as variables for number of continents (Continents) and a variable capturing language heterogeneity (Languages) (see Appendix).

We use three types of regression analyses. First, for the continuous measures of novelty and conventionality we use OLS regression models. Second, we used logistic regression models to test the binary novelty and conventionality measures separately, using the same controls. Third, we use multinomial logistic regression, where the dependent variable has four categories, as described in Figure 4. All models include controls for number of authors, number of references, and field fixed effects. Including the field fixed effect is an important modeling decision, since novelty and conventionality are likely to vary significantly across scientific disciplines.

## 4 - Results

Table 1 shows the descriptive statistics and bivariate correlations for all variables used in the subsequent analyses. All correlations are significant at a p-value below 0.0001, except for the correlation between Novelty and number of authors (Authors), which has a p-value of 0.1087. The correlations show a negative relationship between number of countries (Countries) and Novelty and a positive relationship with Conventionality. The same pattern is observed with the binary versions of these variables (Novelty Bin) and (Conventionality Bin). These correlations suggest the need for further multivariate analysis.

**Table 1 – Descriptive Statistics and Correlations, 2005**

| Variable | N | Mean | Std Dev | 1 | 2 | 3 | 4 | 5 | 6 |
|---|---|---|---|---|---|---|---|---|---|
| **1- Novelty** | 840468 | -56.9 | 274.8 | 1 | | | | | |
| **2- Novelty Bin** | 840468 | 0.35 | 0.5 | 0.179 | 1 | | | | |
| **3- Conventionality** | 840468 | 301.6 | 664.8 | -0.600 | -0.231 | 1 | | | |
| **4- Conventionality Bin** | 840468 | 0.5 | 0.5 | -0.213 | -0.440 | 0.389 | 1 | | |
| **5- Countries** | 841546 | 1.3 | 0.7 | -0.026 | -0.014 | 0.075 | 0.048 | 1 | |
| **6- Authors** | 847512 | 4.4 | 9.0 | 0.002 | 0.028 | 0.013 | 0.012 | 0.390 | 1 |
| **7- References** | 847512 | 22.2 | 16.2 | 0.105 | 0.113 | 0.007 | 0.080 | 0.076 | 0.054 |

Table 1 Notes: All Correlations are significant = p-value<0.0001, except Authors-Novelty; which is non-significant; Data source, WoS.

To illustrate the negative relationship between number of countries and novelty we show the mean tenth percentile z-score by number of countries in Figure 5. The figure shows that the average novelty score decreases slightly as the number of countries grows. This is consistent with the negative correlations observed in Table 2.

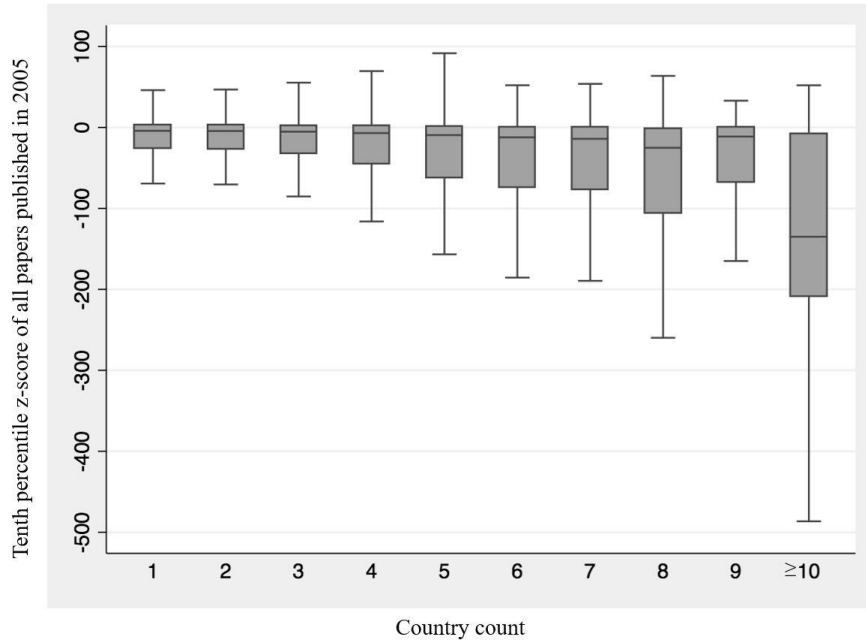

Figure 5. Mean Tenth Percentile Z-Score by Number of Countries, 2005. The z-score has been reverse-coded to be consistent with the analysis as a score for "novelty", showing a decrease in novelty as the number of countries increases.

To illustrate the positive relationship between number of countries and conventionality, Figure 6 shows that the mean of the median z-score increases by number of countries.

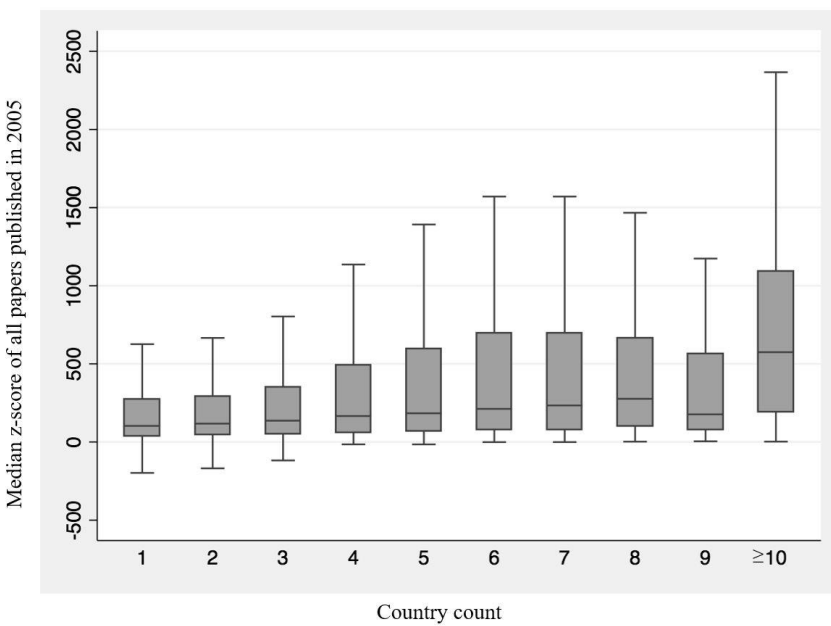

Figure 6. Median z-score by Country Count, 2005. The figure shows an increase in the average conventionality score as the number of countries increases.

Table 2 shows the OLS regression analysis where the dependent variables are novelty and conventionality respectively in the first and second models, measured as continuous z-scores. Each model includes an intercept, Countries, Authors, References, and Field fixed effect. Against expectation, an increase in Countries is negatively associated with Novelty, and positively associated with Conventionality. The opposite is observed for Authors, which are positively associated with Novelty and negatively associated with Conventionality.

**Table 2 – OLS Regression Models, 2005**

|  | Novelty | Conventionality |
|---|---|---|
| **Intercept** | -61.583 | 139 |
|  | (3.89) | (8.337) |
| **Countries** | -2.47 | 17.77 |
|  | (0.475) | (1.018) |
| **Authors** | 0.317 | -1.613 |
|  | (0.035) | (0.074) |
| **References** | 1.614 | 0.173 |
|  | (0.02) | (0.043) |
| **Field** | Fixed | Fixed |
| **N** | 835698 | 835698 |
| **$R^2$** | 0.086 | 0.288 |

Table 2 Notes: All estimates significant, p-value<.0001; standard errors in parentheses; Data source WoS.

Table 3 shows the logistic regression results, where the dependent variables are binary. The results are similar to the regression models in Table 2. Against expectation, Countries is negatively associated with Novelty and positively associated with Conventionality. The opposite is true for Authors, which is positively associated with Novelty and negatively associated with Conventionality. All estimates are significant, p<.0001, so asterisks are not shown for p-values.

**Table 3 – Logistic Regression Models, 2005**

|  | Novelty Bin | Conventionality Bin |
|---|---|---|
| **Intercept** | -1.444 | -0.647 |
|  | (0.036) | (0.03) |
| **Countries** | -0.055 | 0.063 |
|  | (0.004) | (0.004) |
| **Authors** | 0.003 | -0.001 |
|  | (0.0003) | (0.0003) |
| **References** | 0.008 | 0.012 |
|  | (0.0002) | (0.0002) |
| **Field** | Fixed | Fixed |
| **N** | 835698 | 835698 |
| **AIC** | 1017311.7 | 1084349.5 |
| **Wald** | 53534.34 | 60068.87 |

Table 3 Notes: All estimates significant, p<.0001; except Authors in model 1, p=0.0063; Data source, WoS.

Figure 7 illustrates the number of Countries by multinomial category with highest number of articles belonging to Category 3 (LN/HC).

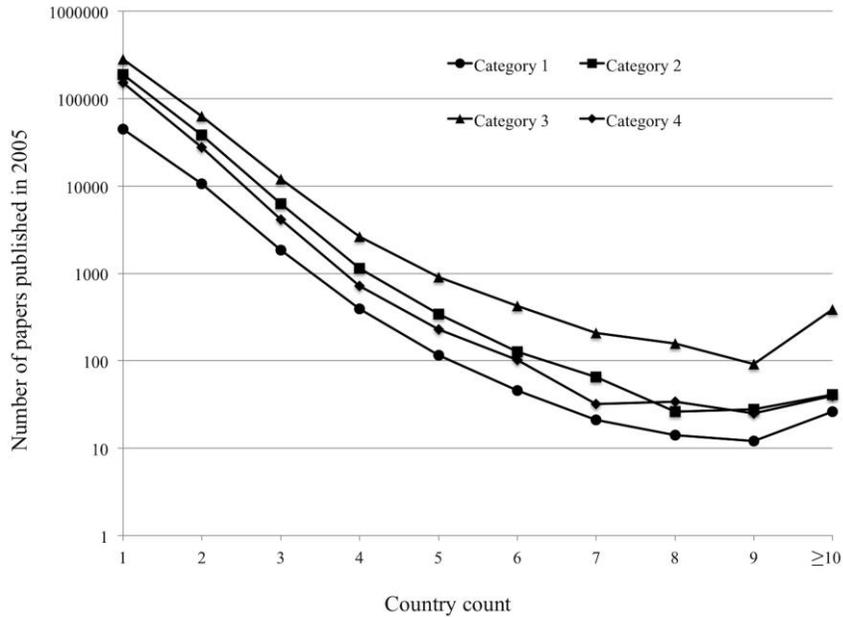

Figure 7. Number of papers by number of countries. For each of the four categories demonstrated in Figure 4, we plot the number of papers published in 2005 as a function of number of countries (country count).

Table 4 shows the results of four multinomial logistic regression models: 1) "All Fields"; 2) "Sciences"; 3) "Social Sciences"; and 4) "Arts and Humanities". Category 4 (LN/LC) is the reference category for all models. For brevity, we do not show the intercept estimates in the models. Against expectation, in the All Fields model (238 fields), Countries is most strongly associated with Category 3 (LN/HC), and Authors is most strongly associated with Category 2 (HN/LC). The Sciences model (165 fields) shows a similar pattern to the All Fields model. The Social Sciences model (42 fields) is different from the Sciences and All Fields models -- the sign on the coefficient for Countries is positive for Category 1 (HN/HC). However, the p-value is .0627, thus the estimate is not significant. The Arts and Humanities model shows a similar pattern, with a positive sign on Category 2 (HN/LC) and 3 (LN/HC) for Countries. However, the Arts and Humanities model also shows a stronger estimate for Countries' association with Category 3 (HN/LC). The Arts and Humanities model does not include the field fixed effect because the model could not converge, and the results could vary depending on the specific sub-field. Also, the Arts and Humanities tend to not have reference lists on articles, which make the novelty/conventionality analysis impossible for many articles.

**Table 4 – Multinomial Logistic Regression Models, 2005**

| | | All Fields | Sciences | Social Sciences | Arts & Humanities |
|---|---|---|---|---|---|
| **Countries** | 1 - (HN, HC) | -0.012 | -0.016 | 0.164 | 0.203 |
| | | (0.0083) | (0.0084) | (0.088) | (0.499) |
| | 2 - (HN, LC) | -0.053*** | -0.057*** | -0.001 | 0.355** |
| | | (0.006) | (0.0061) | (0.039) | (0.131) |
| | 3 - (LN, HC) | 0.025*** | 0.021*** | 0.123*** | 0.535*** |
| | | (0.005) | (0.0056) | (0.028) | (0.113) |
| **Authors** | 1 - (HN, HC) | 0.029*** | 0.03*** | -0.035 | 0.198 |
| | | (0.0013) | (0.0013) | (0.031) | (0.133) |
| | 2 - (HN, LC) | 0.031*** | 0.032*** | 0.072*** | 0.262*** |
| | | (0.0013) | (0.0013) | (0.01) | (0.037) |
| | 3 - (LN, HC) | 0.028*** | 0.029*** | -0.037*** | -0.027 |
| | | (0.0013) | (0.0013) | (0.0086) | (0.037) |
| **References** | 1 - (HN, HC) | 0.036*** | 0.037*** | 0.009** | 0.03 |
| | | (0.0004) | (0.0004) | (0.0033) | (0.015) |
| | 2 - (HN, LC) | 0.029*** | 0.029*** | 0.018*** | 0.035*** |
| | | (0.0003) | (0.0003) | (0.0013) | (0.0042) |
| | 3 - (LN, HC) | 0.03*** | 0.03*** | 0.023*** | 0.028*** |
| | | (0.0003) | (0.0003) | (0.0009) | (0.0035) |
| **Field** | | Fixed | Fixed | Fixed | n/a |
| **N** | | 835698 | 788691 | 39880 | 6069 |
| **AIC** | | 1914595.1 | 1824255 | 76992.56 | 11506.7 |
| **Wald** | | 124820.25 | 107647.96 | 5434.09 | 261.42 |

Table 4 Notes: p<.05*, p<.01**, p<.001***; standard errors in parentheses; Category 4 is reference category; Data source, WoS.

*4.1 Disciplinary Affects*

We also tested whether these results were consistent within specific disciplines, which have been shown to range in terms of citation patterns (Harzing 2010; Ioannidis et al. 2016). Uzzi et al. (2013) also examined the relationship between novelty, conventionality, and disciplines and found that the relationships generally held true, but that differences can be seen when controlling for disciplines. Boyack and Klavans (2014) examined disciplinary effects by taking the top 5% of highly cited articles by discipline, using the article-based discipline-level structure. Their findings suggested that highly cited disciplines have an affect on the relationship between novelty and conventionality. When disciplinary weights are considered, the relationship between novelty/conventionality and citation strength is not as prominent. Although their results confirmed the approximate relationship between categories of novelty and conventionality on citation impact, Boyack and Klavans found that weighting for disciplines influenced the outcome.

We chose fields to reflect a range of referencing behavior, from astrophysics, which has high citing activity, to mathematics, which has low citing activity. We replicated the analysis for six scientific specialties that were analyzed in previous research (Wagner, Whetsell, & Leydesdorff 2017) using WoS data. At this disciplinary level, the results were either non-significant or they supported the main results in Table 4. We expected to find that high-citation fields also show a stronger relationship with Category 1. However, in all six fields, international collaboration was either positively associated with conventionality or negatively associated with novelty, or non-significant. For astrophysics and astronomy, international collaboration was not significant for Category 3 (LN/HC) but there was a significant negative association with Category 2 (HN/LC), suggesting international collaboration tends not to produce novel work. The same patterns held for geoscience/multidisciplinary. For virology, we observed that international collaboration was not significantly associated with any category. For polymers, we found a significant negative association with Category 1 (HN/HC). For agriculture and soil science, we found a significant positive association with Category 3 (LN/HC). Finally, for mathematics, we found negative associations with all categories.

*4.2 Robustness Checks*

We conducted robustness checks, detailed in the Appendix. First, we added four years of data, 2001 up to 2005, and performed identical OLS regressions for each year. We tested the models using different potentially confounding variables. We used number of continents and language as variables, but these move consistently with Countries. We tested models using the natural log of number of Countries and Authors to test for a potential non-linear effect, again with consistent results; Countries has a positive sign for Conventionality and a negative sign for Novelty, and the opposite for Authors. We tested a multinomial logistic regression model where median-z-score calculated by field. These checks produced results that were consistent with the main findings.

**5 - Discussion**

Based upon findings that international collaboration is more highly cited, more diverse, and possibly structured for combinatorial dynamics, we expected to find that it also would be fall into the Uzzi et al. (2013) characterization of articles that are both highly novel and highly conventional, or consisting of "atypical" references, combining diverse parts into new knowledge (Category 1). This is not upheld by the analysis. International collaboration, in aggregate, indicates high conventional measures and low novelty measures for reference combinations. The results were tested for source publications for 2005 in two independent datasets, one using Web of Science used by Uzzi et al. (2013), and one using Scopus by Boyack and Klavans (2014), plus four additional years of WoS data. In addition, using a slightly different method of measuring atypicality, Jian Wang reported similar findings using Web of Science data for the year 2001, in a personal communication[5].

The negative finding against highly atypical referencing behavior in international collaboration, against expectations, led us to consider several possible explanations. We initially considered three possible explanations: 1) the higher transaction costs of international research may mitigate against high novelty; and 2) reliance on information technologies to communicate may reduce interjection of tacit knowledge—which has been shown to contribute to creativity; and 3) the need to rely on English as a common language may limit exchange of information. While these features may hinder creativity, they do not explain why international collaboration is more highly cited than other work. For that, we turned to 4) sociological explanations: an audience effect, which is almost surely at work, and a preferential attachment feature, which is consistent with all the data.

---

[5] July 2018.

*5.1- High transaction costs*

International projects may face higher levels of complexity resulting in higher transactions costs, such as the costs of coordination and communication (e.g., Williamson 1991; Ou, Varriale, & Tsui 2012; Lauto & Valentin 2013). The higher complexity can include the 1) awkwardness of working across time zones; 2) the need to travel periodically long distances to work together; 3) the loss of information due to sub-optimal communication routines; and 4) clashes of management systems (e.g. Leung 2013; Jeong 2014). Any one or combination of these obstacles may suppress otherwise creative or atypical knowledge pairing behavior, as international participants may withhold differences of opinion and defer to a lead author. In this sense, international collaboration may lean towards more hierarchical governance centralized around single or fewer leaders. Different worldviews, nomenclatures, languages, and expectations, can have the effect of slowing the integration of ideas, and may encumber the quality and validity of the results.

Our initial hypothesis suggested that international collaboration might enhance creativity, supported by Stahl et al. (2010) who reported a meta-analysis of 108 multicultural teams and found that cross-cultural connections can lead to increased creativity and satisfaction. Fiore (2008) also found that diverse connections bring creativity to teams. But the results suggest that diversity may also introduce greater transaction costs that could hinder the communication leading to novel or creative ideas. The alternative hypothesis is supported also by Stahl et al. (2010) who found that "cultural diversity leads to process losses through task conflict and decreased social integration…." (p.690) and may create losses through task conflict and decreased social integration.

*5.2 - Reliance on the English Language*

Language differences may be more problematical than cultural differences, especially for those working across countries. Many scholars work in English, even if it is their second language. Lagerström and Andersson (2003) found that members of multilingual teams may not understand questions, assignments or results formulated in English if one or both communicators are not native speakers. Other researchers have also shown the negative effects of language differences on team activities (see Tenzer & Pudelko, 2012). This barrier may reduce opportunities for highly creative discussion that would lead to novel work among collaborators.

*5.3 - Limitations on implicit communications*

Similar to the transaction costs logic discussed above, a further explanation of conventionality could involve the reliance on information-communications technologies (ICTs), which limits the ability to share tacit or implicit knowledge. Groups working at a distance are more likely than physically proximate groups to be using information-communications technologies (ICTs) to exchange information. ICTs favor the transmission of knowledge that can be codified and reduced to data (Roberts 2010). ICTs do not facilitate the exchange of tacit knowledge, defined as the tradition, inherited practices, implied values, and prejudgments held by people involved in a communications process. Polanyi said that tacit knowledge is a crucial part of scientific knowledge (1966). Tacit knowledge is shared through a socialization process; it becomes explicit through externalization (Leonard & Sensiper, 1998). Leonard & Sensiper (1998) wrote: "Researchers stimulating implicit learning found, in fact, that forcing individuals to describe what they thought they understood about implicitly learned processes often resulted in poorer performance than if the individuals were allowed to utilize their tacit knowledge without explicit explanation," (P. 114) meaning learning by doing and imitation are likely more effective ways of transmitting information than 'telling.' Further, they identify three types of tacit knowledge exercised in innovation processes: 1) problem solving, 2) problem finding, and 3) prediction and anticipation, all of which occur within group settings that are "born out of conscious, semiconscious, and unconscious mental sorting, grouping, matching, and melding…" (p. 115). These processes occur at an interpersonal level and are much richer in person than through a written medium. It may be that ICTs cannot substitute for the exchange of tacit knowledge as a critical component of innovation that take place face-to-face. The process of drawing conclusions and making observations will likely occur in a linear fashion, whereas theory suggests that innovation is an iterative process of divergence and convergence in concentric circles (Nonaka, 1994; Leonard & Sensiper, 1998). Distance could make this process difficult or unattainable. This part of the knowledge-creation process may be missing from virtual collaboration, lessening the opportunities for novel outcomes. Leonard and Sensiper (1998) showed that learning by doing and imitation are likely more effective ways of transmitting tacit information than the cognitive process of 'telling' someone about a task or idea. Wuestman et al. (2018) said that "face-to-face interaction between scientists is … necessary to transfer the tacit knowledge that is required to judge the meaning … of a new finding…." (p. 2) ICTs, or the lack of collocation of collaborators enabling tacit information exchange, may inhibit emergence of creative ideas.

More teammates or collaborators may introduce even more complications to the knowledge exchange process. Our research partially supports this proposition in that larger teams of authors are associated with higher novelty and negatively associated with conventionality. This support Wang et al. (2017), in that high-risk taking without shared depth in conventional knowledge has a higher risk of failure, and that large domestic teams produce more high-risk work than large international teams. Thus, face-to-face cooperation and communication may be more important factors for novelty than has been considered in the past. Distance collaborations may not allow the 'bandwidth' for communications needed for creative findings. Sugimoto et al. (2017) showed that researchers who are more mobile have about 40% higher citation rates than non-mobile ones, again supporting the need for face-to-face work. Similarly, Wagner, et al. (2018) found that more 'open' countries—those welcoming newcomers, sending researchers to work abroad, and coauthoring internationally—had higher citation impacts than other countries. These findings suggest that people working side-by-side may have a better outcome than those working virtually.

*5.4 - Audience effect of international coauthorship*

Transaction costs, communication obstacles, and cultural differences are possible explanations for the high conventionality of international collaboration; however, these factors do not explain why international coauthorships are more highly cited. One would expect that less creative work would be less cited than other work. This spurred us to query other literature that has suggested that international collaborative articles are benefitting from an agglomeration effect, an audience factor (Zitt & Small, 2008), or a "bonus effect" (Kato & Ando, 2013). Network studies show that when performance is difficult to measure – as it is with scientific or other scholarly pursuits – network connections, more than performance, are closely tied to success (Guimera et al., 2005). Schmoch and Schubert (2008) suggested that international articles are more highly cited because their readership community is larger. Börner et al. (2006) found that citations tended to occur in geographically proximate places first before diffusing outward. People may cite others close-by at first, so articles with multiple, dispersed addresses, may have broader citation possibilities. In this case, international collaboration might be a "force multiplier" for the numbers of people in the network of readers and citers (Lancho-Barrantes et al., 2012). More countries and more authors could mean more citations, whether or not the work is more creative. While an audience effect is a more satisfying explanation for the greater attention to international work than the three discussed above, one more part of the puzzle still defies explanation: why citation strength for international work persists over long periods of time. It could be argued that an audience effect could help citations in the short term, but not in the longer term. One would expect the audience effect to fade quickly.

To explore this question, we turned to a sociological explanation. The findings fully support a social dynamic related to gaining citations through a process of preferential attachment, where reputation (rather than novelty) is the operative dynamic. Wagner and Leydesdorff (2005) showed that international collaborative networks are evolving through a pattern of preferential attachment—where researchers seek to attach themselves to more reputed collaborators in order to enhance their own reputation. The process is one of aspirational collaboration with a person of high or higher reputation and resources. In the process of making an aspirational connection, one enhances one's own reputation-by-affiliation, and thereby gains attention to one's work. Moreover, a reputation effect has the further virtue of explaining why international collaboration is growing at a high rate. More practitioners worldwide are scrambling for recognition and reward, even as resources remain flat. Seeking attention on a broad 'stage' (offered at the international level) reaps rewards at the disciplinary and institutional levels.

This explanation resonates with Whitley (2000) who observed that scholarship contains a tension distinguished by a commitment to produce novelty, on one hand, and the need for reputation, on the other. Indeed, Whitley says: "…the emphasis of the system on fame and fortune following from convincing a large number of influential colleagues of the importance of one's work ensures general adherence to current procedural norms…." (p. 23), suggesting that strong reputation may even mitigate against novelty. Once reputation is gained, maintaining it can be done by adding incremental quanta of knowledge that do not risk disciplinary controversy, nor risk falling through the cracks between disciplines. In addition, the academic reward system does not always recognize value in interdisciplinary work such as those articles that may contain 'atypical' referencing pairs. Goring et al. (2014) point out that "….even within interdisciplinary organizations such as LTER [Long-Term Ecological Research], early career researchers still largely engage in projects within the single discipline" because of concern about adverse promotion and tenure impacts.

The citation environment at the international level, then, could lean towards rewarding reputation as much or more than novelty, in part because the number of ties are higher and the reach to others is less specialized. It may also be the case, as Reagans and McElviy (2003) showed, that network range and social cohesion ease knowledge transfer. Thus, it is consistent with the findings in this article that citation strength of international work reflects network strength as much or more than quality or novelty. Indeed, within networks, reputation is a core driver of cohesion. Reputation is "sticky" and it can persist even beyond the point where an author's work is not particularly novel. The suggestion here is that international collaborations are notable because of the reputation of more elite scholars: others seek to attach to them, and as they do, reputation is enhanced in a virtuous cycle. The resulting output is cited because of the reputations of the authors. This could account for both a negative finding around novelty, but higher citation counts in both the short and the long term. This needs more exploration, especially within individual disciplines.

## 6 - Policy Implications & Conclusions

Policymakers and program managers have tended to favor international collaboration in funding decisions (He, 2009). As more elite scientists participate in international collaboration, funding has followed. International projects claim substantial shares of national research and development budgets. No doubt, many international projects are productive and useful. However, the findings here suggest that there may be losses of creativity and novelty associated with distance.

International collaboration is undertaken for many reasons, as discussed by Beaver and Rosen (1978), and Katz and Martin (1997). In some cases, such as "big science" collaborations, large-scale equipment is needed. The infrastructure costs of research preclude any single country from undertaking it. Huge up-front investments are required to research particle physics, astrophysics, or geophysics. Other work, such virology, can require visits with on-the-ground researchers facing a specific outbreak, like the Zika virus, springing up in a unique location. Still other works can only proceed by sharing data or samples. These types of projects require collaborations.

Other collaborations, such as those seeking an audience effect or reputation enhancement, may not have the characteristics of research that engenders highly creative work. In these cases, attention to facilitating enriched communications may be needed. This may mean ensuring that projects have enough budget to support frequent face-to-face meetings. It may also mean discouraging very large teaming arrangements (especially those that do not center around large equipment), as these seem to be antithetical to novelty/conventionality sweet spot communication identified by Uzzi et al. (2013). Further, it may mean, in some cases, finding alternatives to international collaboration when research is highly exploratory and where a good deal of cross-disciplinary discussion is required to move the field forward. An explicit plan to nurture research in this sweet spot may be needed for internationally collaborative projects.

**Appendix - Robustness Checks & Tests**

We conducted the following robustness checks: (1) logistic regression and multinomial regression models using Scopus data for 2005 (Table A.1; A.3); (2) OLS regressions for four additional years of data (Table A.3); (3) Fixed-effects models using different potentially confounding variables, using number of Continents and a language variable (Table A.4); (4) OLS models using the natural log of number of countries and authors to test for a potential non-linear effect (Table A.5); (5) a multinomial logistic regression model where the median z-score, used to create the nominal variable, was first created by field to account for within field variation (Table A.6); (6) a replication test of the effect of international collaboration on citation strength, showing a strong effect (Table A.7). All of the robustness checks were consistent with the main results.

The results of the logistic regression analysis on the Scopus data are presented in Table A.1. Consistent with the primary results, the number of countries affiliated is negatively and significantly related to novelty and is positively and significantly related to conventionality.

**Table A.1 - Logistic Regression, Scopus Data, 2005**

|            | Novelty Bin | Conventionality Bin |
|------------|-------------|---------------------|
| **Intercept** | -0.2393     | -0.1201             |
|            | (0.00452)   | (0.00445)           |
| **Countries** | -0.0796     | 0.1045              |
|            | (0.00326)   | (0.0032)            |
| **N**      | 1048575     | 1048575             |
| **AIC**    | 1423358     | 1452500.7           |
| **Wald**   | 595.3256    | 1065.4101           |

Table A.1 notes: all estimates significant, p<.0001, standard errors in parentheses

Table A.2 shows the results of the multinomial logistic regression on the Scopus data. Category 4 is the reference. Consistent with the Table 4, The number of countries is most strongly associated with Category 3. Number of countries is also negatively associated with Category 2.

**Table A.2 - Multinomial Logistic Regression, Scopus Data, 2005**

|  | 1 - (HN, HC) | 2 - (HN, LC) | 3 - (LN, HC) |
|---|---|---|---|
| **Intercept** | -0.7255 | 0.7193 | 0.7986 |
|  | (0.00911) | (0.00687) | (0.0064) |
| **Countries** | 0.0519 | -0.0375 | 0.0858 |
|  | (0.0065) | (0.00502) | (0.0046) |
| **N** | 1048575 | | |
| **AIC** | 2604447.8 | | |
| **Wald** | 1152.6412 | | |

Table A.2 notes. All estimates significant, $p<.0001$, standard errors in parentheses

Next, Table A.3 shows the results of the analysis by year to demonstrate the consistency of the results across time. The results are consistent with the primary results in the full article. Countries is positively associated with conventionality, and negatively associated with novelty. The opposite is true for authors.

**Table A.3 – OLS Regression Analysis by Year, 2001-2004**

| | Conven. | Conven. | Conven. | Conven. | Novelty | Novelty | Novelty | Novelty |
|---|---|---|---|---|---|---|---|---|
| **Intercept** | 144.972*** | 132.869*** | 142.942*** | 147.079*** | -73.16*** | -65.151*** | -70.427*** | -68.999*** |
| | (8.537) | (7.876) | (8.814) | (8.633) | (4.147) | (3.937) | (4.264) | (4.15) |
| **Countries** | 11.775*** | 15.512*** | 17.73*** | 14.183*** | -1.297* | -1.558** | -2.105*** | -0.86 |
| | (1.14) | (1.032) | (1.101) | (1.043) | (0.554) | (0.516) | (0.533) | (0.502) |
| **Authors** | -1.416*** | -1.432*** | -1.342*** | -1.507*** | 0.256*** | 0.303*** | 0.362*** | 0.282*** |
| | (0.101) | (0.095) | (0.105) | (0.08) | (0.049) | (0.048) | (0.051) | (0.038) |
| **References** | 0.251* | 0.087*** | 0.251*** | 0.293*** | 1.894*** | 1.764*** | 1.836*** | 1.742*** |
| | (0.046) | (0.042) | (0.045) | (0.045) | (0.023) | (0.021) | (0.022) | (0.021) |
| **Year** | 2001 | 2002 | 2003 | 2004 | 2001 | 2002 | 2003 | 2004 |
| **Field** | Fixed | Fixed | Fixed | Fixed | Fixed | Fixed | Fixed | Fixed |
| **R²** | 0.272 | 0.272 | 0.311 | 0.316 | 0.093 | 0.096 | 0.112 | 0.1 |
| **N** | 698599 | 718148 | 754690 | 792861 | 698599 | 718148 | 754690 | 792861 |

Table A.3 Notes. p<.05*, p<.01**, p<.001***, standard errors in parentheses

Next, Table A.4 shows the analysis of alternative variables, including number of continents, and a variable for linguistic composition. Because a straightforward addition of languages is complicated by the fact that many individual countries have multiple languages, we calculated whether the article included no-common languages, a value of one, or at least one common language, zero. We employed a two-way fixed effects approach, where the field and the year are included as fixed classification variables in the regression, while using the full five years of data. However, because of high variance inflation, between continents and languages, we chose to treat each variable in a separate model.

**Table A.4 – Two-Way Fixed Effects Regression, Continents & Languages, 2001-2005**

|  | Conven. | Conven. | Novelty | Novelty |
|---|---|---|---|---|
| **Intercept** | 138.609 | 164.163 | -67.17 | -69.254 |
|  | (3.891) | (3.798) | (1.871) | (1.827) |
| **Continents** | 24.869 |  | -1.63 |  |
|  | (0.763) |  | (0.367) |  |
| **Languages** |  | 13.118 |  | 1.487 |
|  |  | (0.724) |  | (0.348) |
| **Authors** | -1.159 | -1.046 | 0.266 | 0.247 |
|  | (0.037) | (0.037) | (0.018) | (0.018) |
| **References** | 0.209 | 0.223 | 1.762 | 1.757 |
|  | (0.02) | (0.02) | (0.01) | (0.01) |
| **Year** | Fixed | Fixed | Fixed | Fixed |
| **Field** | Fixed | Fixed | Fixed | Fixed |
| **$R^2$** | 0.29 | 0.29 | 0.096 | 0.096 |
| **N** | 3799996 | 3799996 | 3799996 | 3799996 |

Table A.4 notes. All estimates significant, p<.0001, standard errors in parentheses

The Table A. 5 shows log transformed variables for countries and authors. The results are consistent with the primary results. Log of countries is positively related to conventionality and negatively related to novelty. The opposite is true of log of authors.

**Table A.5 – OLS Regression, Log-Transformed Variables, 2005**

|  | Conventionality | Novelty |
|---|---|---|
| **Intercept** | 176.907 | -72.792 |
|  | (8.305) | (3.875) |
| **Log_Countries** | 35.194 | -6.384 |
|  | (1.934) | (0.902) |
| **Log_Authors** | -28.776 | 11.139 |
|  | (1.127) | (0.526) |
| **References** | 0.2 | 1.601 |
|  | (0.044) | (0.02) |
| **$R^2$** | 0.289 | 0.086 |
| **N** | 835698 | 835698 |

Table A.5 notes. All estimates significant, p<.0001, standard errors in parentheses

Next, Table A.6 shows the results of the multinomial logistic regression analysis where the median z-score, is first calculated within individual fields, rather than across all fields, before binarizing and compiling into a nominal four category variable. The results are very similar to Table 4. However, the effect of countries on category 1 is now positive, but not significant. The estimate on category 2 remains negative and significant, and the association with category 3 remains positive and significant.

**Table A.6 – Multinomial Logistic Regression, Conventionality by Field, 2005**

|  |  | All Fields |
|---|---|---|
| **Countries** | 1 - (HN, HC) | 0.015 |
|  |  | (0.008) |
|  | 2 - (HN, LC) | -0.044*** |
|  |  | (0.006) |
|  | 3 - (LN, HC) | 0.036*** |
|  |  | (0.005) |
| **Author** | 1 - (HN, HC) | 0.003*** |
|  |  | (0.0005) |
|  | 2 - (HN, LC) | 0.003*** |
|  |  | (0.0004) |
|  | 3 - (LN, HC) | -0.001*** |
|  |  | (0.0004) |
| **References** | 1 - (HN, HC) | 0.039*** |
|  |  | (0.0003) |
|  | 2 - (HN, LC) | 0.029*** |
|  |  | (0.0003) |
|  | 3 - (LN, HC) | 0.03*** |
|  |  | (0.0003) |
| **Field** |  | FIXED |
| **N** |  | 834998 |
| **AIC** |  | 1975932.1 |
| **Wald** |  | 80123.21 |

Table A.6 notes. p<.05*, p<.01**, p<.001***, standard errors in parentheses

Since the argumentation regarding a potential audience effect relies on the citation strength of international collaboration, Table A.7 shows the citation analysis. The analysis shows a strong significant estimate of 4.15 for Countries on Citations.

**Table A.7 – International Collaboration and Citations, 2005**

|  | Citations |
|---|---|
| **Intercept** | -6.224 |
|  | (0.587) |
| **Countries** | 4.154 |
|  | (0.072) |
| **Authors** | 0.146 |
|  | (0.005) |
| **References** | 0.504 |
|  | (0.003) |
| **Field** | Fixed |
| **R$^2$** | 0.093015 |
| **N** | 841546 |

Table A.7 notes. All est. sig. p<.0001, standard errors in parentheses